\documentclass[11pt,,showpacs]{article}  

\usepackage{amssymb,amstext,amsmath,amsthm}
\usepackage{MnSymbol}
\usepackage[dvips]{graphicx}
\usepackage{latexsym}
\usepackage{psfrag}
\usepackage{amsfonts}
\usepackage{bm} 
\usepackage{amssymb}
\usepackage{bbm}
\usepackage{color}
\usepackage{tikz}  
\usepackage{epigraph}
\usepackage{graphicx}
\usepackage{authblk}

\newcommand{\be}{\begin{equation}}
\newcommand{\ee}{\end{equation}}
\newcommand{\barray}{\begin{array}}
\newcommand{\earray}{\end{array}}
\newcommand{\bea}{\begin{eqnarray}}
\newcommand{\eea}{\end{eqnarray}}
\newcommand{\bs}{\begin{subequations}}
\newcommand{\es}{\end{subequations}}
\newcommand{\balign}{\begin{align}}
\newcommand{\ealign}{\end{align}}
\newcommand{\equ}{\begin{equation}}
\newcommand{\nequ}{\end{equation}}
\newcommand{\eqa}{\begin{eqnarray}}
\newcommand{\neqa}{\end{eqnarray}}
\def\bphi{\bm\varphi}
\def\wc{\widetilde{C}}

\def\nn{\nonumber}

\def\e{\epsilon}

\newcommand{\rd}{\mathrm{d}}

\def\maps{\colon}

\def\to{\rightarrow}

\def\bs{\bar{s}}

\usepackage{bbm}

\newcommand{\SU}{\mathrm{SU}}

\def\vphi{\varphi}
\def\cB{{\cal B}}
\def\cF{{\cal F}}

\def\cA{{\cal A}}
\def\cE{{\cal E}}
\def\cG{{\cal G}}
\newcommand{\Tr}{\mathrm{Tr}}

\def\1{\mathbb{1}}

\begin{document}

\title{Weighting bubbles in  group field theory}
\author[1]{Aristide Baratin\thanks{abaratin@uwaterloo.ca}}
\author[2]{Laurent Freidel\thanks{lfreidel@perimeterinstitute.ca}}
\author[2,3]{Razvan Gurau\thanks{rgurau@cpht.polytechnique.fr}}
\affil[1]{\small \it Department of Applied Mathematics, University of Waterloo, Waterloo, Ontario, Canada}
\affil[2]{\it Perimeter Institute for Theoretical Physics, Waterloo, Ontario, Canada}
\affil[3]{\it CPHT Ecole Polytechnique, UMR 7644 CNRS, Palaiseau, France}

\date{\vspace{-5ex}}

\maketitle

\begin{abstract}
Group field theories (GFT) are higher dimensional generalizations of matrix models whose Feynman diagrams are dual to triangulations. 
Here we propose a modification of GFT models 
that includes extra field indices keeping track of the  bubbles of the graphs in the Feynman evaluations. In dimension three, our model exhibits new symmetries,  interpreted as the action of the vertex translations of the triangulation. The extra field indices have an elegant algebraic interpretation: they encode the structure of a semi-simple algebra. Remarkably, when the algebra is chosen to be associative, 
the new structure contributes a topological invariant from each bubble of the graph to the Feynman amplitudes. 
\end{abstract}

\epigraph{\it ``Double, double toil and trouble \\
Fire burn, and cauldron bubble.''}{\textup{W. Shakespeare,}  \it{Macbeth}}

\section{Introduction}

Group field theories \cite{overview}  (GFT) generalize matrix models \cite{mm} to higher dimensions and provide an elegant field theoretic 
formulation of spin foam models \cite{overview,tenquestions} of quantum gravity.
Their Feynman expansion generates fat graphs, hence having not only vertices, edges and faces, but also higher dimensional cells coined `bubbles'. Just as each graph in matrix models represents by duality a triangulated Riemann surface, each graph in a $D$-dimensional group field theory can be thought of as representing
a $D$-dimensional triangulated (pseudo)-manifold. Since the introduction of the colored \cite{color1} models and their $1/N$ 
expansion \cite{Raz2}, progress has been made in the analytical study of GFTs. Their leading order behaviour has been studied \cite{melogft}
and some renormalizable GFT models \cite{su2} have been introduced.

Bubbles play a key role in the structure of the Feynman amplitudes. In fact, due to the gauge invariance and the specific ultra-local nature of the couplings in GFT models, the  amplitudes contain bubble divergences, analogous to the loop divergences in ordinary quantum field theory. In the context of spin foam models, it has been argued  -- and illustrated explicitly in dimension three -- that such divergences result from a discrete residual action of the diffeomorphism group, acting as translations on the vertices of the triangulation \cite{LaurentDavid}. The first motivation of the present paper, following the line of thoughts developed in \cite{Baratin}, is to investigate ways to encode the action of vertex translations as a symmetry of the group field theory. In fact, as recognized in \cite{Baratin},  GFT already possesses  such a symmetry in the case of an abelian gauge group. 
Building upon a dual formulation of GFT as a non-commutative field theory of Moyal-type \cite{Baratin-Oriti, Flo}, 
the proposal of \cite{Baratin} in the non-abelian case is to implement this symmetry as a deformed symmetry,  by promoting the fields into braided  fields -- obeying a braided statistics \cite{Oeckl}. The difficulty with such a proposal\footnote{For analogous proposals in different contexts, see for example \cite{FloEtera} or \cite{Balachandran}.} is that it immediately brings us outside the realm of standard field theory, where fields are real or complex valued. In particular, we loose control on the measure used to define the functional integral; this measure can be defined formally \cite{Oeckl, LivineF, LivineF1, Sasai} but not explicitly. Last but not least, no analogue of Noether charges or Ward identities exist  just yet in this context. 
 
Interesting as it may be to try to gain a better understanding of braided (group) field theory and to extend the notion of what is meant by a symmetry,
here we would like to take a more standard route. Our aim is to investigate the construction of a GFT model invariant 
under a standard symmetry expressing the translation of vertices for non-abelian groups. 

The idea we follow is simple: we just extend the definition of the fields to include extra indices labeling the vertices of the tetrahedron patterned by the interaction. Making this explicit in dimension three leads us to a modification of the Boulatov model \cite{Boulatov} giving  
rise to an extra contribution to the Feynman amplitudes from each bubble of the graph.  As we  will show, our modified model has an elegant algebraic formulation: the extra field labels encode the structure of a semi-simple algebra.  The GFT action takes the form of a trace invariant in this algebra, which makes explicit the presence of new (unitary) symmetries expressing vertex translation invariance. Moreover, remarkably, when the algebra is chosen to be associative, the new bubble contribution defines a topological invariant characterizing the topology of the bubble.  

We begin in Section 2 by recalling the issue of vertex translation symmetry in the context of three dimensional colored GFT. In Section 3 we define our new model and give its algebraic formulation. We compute the Feynman amplitudes in Section 4 and discuss in detail various aspects and consequences of their structure.   
We conclude in  Section 5 with some directions for future work.

\section{GFT and translational symmetry}

The archetypal group field theory in three dimensions is the Boulatov model, whose Feynman graph expansion generates a topological $BF$
evaluation of each Feynman diagram \cite{Boulatov}. The dynamical variable is a complex field  $ \vphi(g_{1}, g_{2}, g_{3})$ on a product of 
three copies of a group $G$. The {\sl colored} model \cite{color1,color2} 
depends on four such fields $\vphi_{i}(g_{1},g_{2},g_{3})$, $i=1,\cdots, 4$; the label $i$ is the color of the field. The fields are assumed 
to satisfy the gauge symmetry:
\be\label{gauge1}
\vphi_{i}(g_{1}g, g_{2}g, g_{3}g) = \vphi_{i}(g_{1}, g_{2}, g_{3}), \quad \forall g \in G.
\ee
The kinetic term is given by a purely ultra-local coupling
\be
K\equiv \sum_{i=1}^4\int \rd g_{1}  \rd g_{2}\rd g_{3} \, \bar{\vphi}_{i}(g_{1}, g_{2}, g_{3})\vphi_{i}(g_{1}, g_{2}, g_{3}).
\ee
where $\bar{\vphi}_{i}$ denotes the complex conjugate of $\vphi_i$.  
The interaction is given by $ \lambda(V + \bar{V})$ where $\lambda$ is the coupling constant and $V$ is the potential given by
\be
V\equiv \int \prod_{i<j}\rd g_{ij} \, \vphi_{1}(g_{14}, g_{13}, g_{12})\vphi_{2}(g_{21}, g_{24}, g_{23})
\vphi_{3}(g_{32}, g_{31}, g_{34})\vphi_{4}(g_{43}, g_{42}, g_{41})
\ee
where we set $ g_{ji}= g_{ji}$ in the integrand. Given $(i,j,k,l)$ all distinct and $(j,k,l)$ cyclically ordered, we may view the three arguments $(g_{ij},g_{ik},g_{il})$ of the fields $\vphi_{i}$ as representing the three edges $(ij),(ik),(il)$ of a triangle:
\[
\begin{array}{c} 
\begin{tikzpicture}[scale=0.7]
\draw (0,0) -- (2,3) -- (4, 0) --cycle;
\draw[dashed, thick]  (2, 1)-- (1.09, 1.62);
\draw[dashed,thick] (2,1) -- (2.91, 1.62);
\draw[dashed,thick] (2,1)--(2,0);
\filldraw[black] (2,1) circle (1.5pt);
\path (0.6,1.8) node {\footnotesize $(ij)$};
\path (3.4,1.8) node {\footnotesize$(ik)$};
\path (2,-0.4) node {\footnotesize $(il)$};
\path (2.25,0.9) node {\footnotesize$i$};
\end{tikzpicture}
\end{array}
\]
In this simplicial picture, the identification of group elements in the quartic interaction $V$ follows the combinatorial pattern of a tetrahedron. 
Each triangle of the tetrahedron is thus labelled by a color $i$ and $(ij)$ is the edge common to the two triangles $i$ and $j$.


\subsection{Abelian symmetry}

The tetrahedral interaction  described above is special. As first recognized in \cite{Baratin}, it possesses extra symmetries when $G$  
is an abelian group, which we assume to be U(1) for concreteness.  They are four such symmetries, each of which is generated by a group
character, i.e a complex function $\chi$ on $G$ such that  $\chi(g)\chi(h)=\chi(gh)$ and $\bar{\chi}(g)=\chi(g^{-1})$. They  are interpreted 
as translational symmetries acting at the {\sl vertices} of the tetrahedron. 

In the following,  we will denote the vertex opposite to the triangle ${1}$ (respectively 2,3,{4})  by the capital letter $A$ (respectively $B,C,D$). 
Thus, the vertex $A$, to which corresponds a triple of colors $(234)$,  lies at the intersection of the three triangles $2, 3, 4$ and is common to 
the edges $(23)$, $(34)$ and $(42)$. The translation $T_A$ of the vertex $A$, generated by the U(1) character $\chi_A$, acts on the fields as follows:
\bea \label{TA}
T_{A}(\vphi_{1})(g_{14}, g_{13}, g_{12})&=& \vphi_{1}(g_{14}, g_{13}, g_{12}), \nn\\
T_{A} (\vphi_{2})(g_{21}, g_{24}, g_{23}) &=& \chi_A(g_{24}^{-1}g_{23}) \vphi_{2}(g_{21}, g_{24}, g_{23}),\nn \\
T_{A} (\vphi_{3})(g_{32}, g_{31}, g_{34})&=&  \chi_A(g_{32}^{-1}g_{34}) \vphi_{3}(g_{32}, g_{31}, g_{34}),\nn\\
T_{A} (\vphi_{4})(g_{43}, g_{42}, g_{41})&=& \chi_A(g_{43}^{-1}g_{42}) \vphi_{4} (g_{43}, g_{42}, g_{41}).
\eea
Thus, given a triangle $i$ having $A$ as one of its vertex, $T_A$ acts on $\vphi_i$ by multiplication by $\chi(g^{-1}g')$, 
where $g,g'$ are the two group elements associated with the two edges of the triangle that touches $A$. With this understanding it is straightforward 
to write the action of the translations $T_B, T_C, T_D$ of the three other vertices. 

A key property of the transformation (\ref{TA})  is that it respects the gauge symmetry (\ref{gauge1}): 
\be
T_A(\vphi_{i})(g_{ij}g,g_{ik}g,g_{il}g) = T_A(\vphi_{i})(g_{ij},g_{ik},g_{il}).
\ee
It is clear also that it is a symmetry of the kinetic and interaction terms,  thanks respectively to the conjugation and the multiplicative property
of the character. As shown in \cite{Baratin}, the existence of this symmetry is related to the topological translational symmetry of the corresponding 
spin foam model. It is important to note that the very possibility to implement this symmetry as a field transformation is intimately tied to the field
coloring, which allows to distinguish between the different vertices of the tetrahedron. 

In order to generalize this symmetry to the non-abelian case,  it will be convenient to write the same symmetry transformation under the alternate form:
\bea 
T_{A}(\vphi_{1})(g_{14}, g_{13}, g_{12})&=& \vphi_{1}(g_{14}, g_{13}, g_{12}), \nn\\
T_{A} (\vphi_{2})(g_{21}, g_{24}, g_{23}) &=& \chi_A(g_{24}^{-1}) \vphi_{2}(g_{21}, g_{24}, g_{23}) \chi_A(g_{23}), \nn\\
T_{A} (\vphi_{3})(g_{32}, g_{31}, g_{34})&=&  \chi_A(g_{32}^{-1}) \vphi_{3}(g_{32}, g_{31}, g_{34})\chi_A(g_{34}),\nn\\
T_{A} (\vphi_{4})(g_{43}, g_{42}, g_{41})&=& \chi_A(g_{43}^{-1}) \vphi_{4} (g_{43}, g_{42}, g_{41})\chi_A(g_{42}).\label{transab}
\eea
obtained from (\ref{TA}) by using the multiplicative property of the character. Note also that for the transformation (\ref{transab}) to be a symmetry,
the character arguments do not need to be the arguments of the fields; we may in fact consider the more general transformations:
\bea 
\tilde{T}_{A}(\vphi_{1})(g_{14}, g_{13}, g_{12})&=& \vphi_{1}(g_{14}, g_{13}, g_{12}), \nn\\
\tilde{T}_{A}(\vphi_{2})(g_{21}, g_{24}, g_{23}) &=& U_{24}^{-1} \vphi_{2}(g_{21}, g_{24}, g_{23}) U_{23}, \nn\\
\tilde{T}_{A}(\vphi_{3})(g_{32}, g_{31}, g_{34})&=&  U_{32}^{-1} \vphi_{3}(g_{32}, g_{31}, g_{34})U_{34},\nn\\
\tilde{T}_{A}(\vphi_{4})(g_{43}, g_{42}, g_{41})&=& U_{43}^{-1} \vphi_{4} (g_{43}, g_{42}, g_{41})U_{42}.\label{transab2}
\eea
where $U_{ij} = U_{ji}$, $\bar{U}_{ij}= U_{ij}^{-1}$ are arbitrary U(1) elements.

This construction however, which relies on the existence of a complex valued character,  works only for abelian group.  This is clearly 
disappointing since the quantum gravity models always rely on the use of a non-commutative (Lorentz) group.
In the work \cite{Baratin}, it is proposed to implement this symmetry in the non-commutative case as resulting from the action of a (quantum) deformation\footnote{When  $G=\SU(2)$, the relevant quantum symmetry group is the (translational part of the) Drinfeld double DSU(2). The role of DSU(2) in 3d quantum gravity with vanishing cosmological constant has been known for some time \cite{BaisMuller, PRII}, so it should not be surprising to see it naturally show up in the context of GFT (see also \cite{Livine-Girelli}).}  of the translation group. However the difficulty with this proposal is that in order for  the fields $\vphi_{i}$ to carry a representation of the quantum group, they must be promoted to braided fields -- obeying a braided statistics. As mentioned in the introduction, this brings us outside the realm of standard field theory; progress is very challenging in this context. 
Here we would like to take a more standard route and remain within the usual field theory framework.

\section{A model with more indices}

To be able to extend the translational symmetry to non-abelian groups, the idea is to modify the definition of the colored Boulatov model by adding extra indices to the fields labeling the vertices of the tetrahedron patterned by the interaction. These extra indices will allow us to keep track of the bubbles forming in the Feynman graph expansion. 

\subsection{Action}

We thus consider an extended model, described by colored fields $\vphi_{i}^{ABC}(g_1, g_2, g_3)$ carrying three extra indices $A,B,C$ running over 
a finite set $I$. The  fields are still assumed to be invariant under a global shift of their arguments:
\be
\vphi_{i}^{ABC}(g_{1}g,g_{2}g, g_{3}g) = 
\vphi_{i}^{ABC}(g_{1},g_{2}, g_{3}) \; . \label{gauge2}
\ee
The GFT action that we propose has a kinetic term of the form:
\be\label{kindices}
K =  \sum_{i=1}^4 \int\rd g_{1}\rd g_{2}\rd g_{3} \,\, h_{\bar A A} h_{\bar B B} h_{\bar C C} 
\, \bar{\vphi}^{\bar A\bar B\bar C}_{i}(g_{1}, g_{2}, g_{3}) \vphi^{ABC}_{i}(g_{1}, g_{2}, g_{3})
\ee
where repeated indices $A, \bar A...$ are implicitly summed over. The kinetic term depends on a rank 2 
tensor $h_{AB}$; we will choose this tensor to be hermitian $h_{AB} = \overline{h_{BA}}$ and non-degenerate, so 
that the kinetic term is real and invertible.  The interaction is given by $ \lambda(V + \bar{V})$,  where $\lambda$ is the coupling constant and:
\bea \label{intdices}
V =  &&    \int \prod \rd g_{ij}  \prod_{i<j} \delta(g_{ij} g_{ji}^{-1})  \, \, C_{ A_2 A_3 A_4  } 
        C_{ B_1 B_3 B_4  } C_{ C_1 C_2 C_4  } C_{ D_1 D_2 D_3  } \\
    &&\hspace{-1cm}  \vphi_1^{B_1 C_1 D_1 }(g_{14}, g_{13}, g_{12}) \vphi_2^{A_2 C_2 D_2 }(g_{21}, g_{24}, g_{23}) 
     \vphi_3^{A_3 B_3 D_3 }(g_{32}, g_{31}, g_{34} )  
     \vphi_4^{A_4 B_4 C_4 }(g_{43}, g_{42}, g_{41} ) \; . \nonumber
   \eea
We see that the new field indices are contracted by means  of a rank 3 tensor $C$.  In the simplicial picture where the arguments of each field are associated to the three edges of a triangle, the additional indices are associated to the vertices:
\[
\begin{array}{c} 
\begin{tikzpicture}[scale=0.6]
\draw (0,0) -- (2,3) -- (4, 0) --cycle;
\draw[dashed, thick]  (2, 1)-- (1.09, 1.62);
\draw[dashed,thick] (2,1) -- (2.91, 1.62);
\draw[dashed,thick] (2,1)--(2,0);
\draw[dotted, thick] (2,1) --(2,3);
\draw[dotted, thick] (2,1) --(0,0);
\draw[dotted, thick] (2,1) --(4,0);
\filldraw[black] (2,1) circle (1.5pt);
\path (0.6,1.8) node {\tiny $(14)$};
\path (3.4,1.8) node {\tiny $(13)$};
\path (2,-0.4) node {\tiny $(12)$};
\path (2.4,1) node {\tiny $1$};
\path(2,3.3) node {\tiny $B_1$};
\path (-0.3, -0.1) node {\tiny $C_1$};
\path (4.4, -0.1) node {\tiny $D_1$};
\end{tikzpicture}
\end{array}
\quad 
\begin{array}{c} 
\begin{tikzpicture}[scale=0.6]
\draw (0,0) -- (2,3) -- (4, 0) --cycle;
\draw[dashed, thick]  (2, 1)-- (1.09, 1.62);
\draw[dashed,thick] (2,1) -- (2.91, 1.62);
\draw[dashed,thick] (2,1)--(2,0);
\draw[dotted, thick] (2,1) --(2,3);
\draw[dotted, thick] (2,1) --(0,0);
\draw[dotted, thick] (2,1) --(4,0);
\filldraw[black] (2,1) circle (1.5pt);
\path (0.6,1.8) node {\tiny $(21)$};
\path (3.4,1.8) node {\tiny $(24)$};
\path (2,-0.4) node {\tiny $(23)$};
\path (2.4,1) node {\tiny $2$};
\path(2,3.3) node {\tiny $C_2$};
\path (-0.3, -0.1) node {\tiny $D_2$};
\path (4.4, -0.1) node {\tiny $A_2$};
\end{tikzpicture}
\end{array}
\quad
\begin{array}{c} 
\begin{tikzpicture}[scale=0.6]
\draw (0,0) -- (2,3) -- (4, 0) --cycle;
\draw[dashed, thick]  (2, 1)-- (1.09, 1.62);
\draw[dashed,thick] (2,1) -- (2.91, 1.62);
\draw[dashed,thick] (2,1)--(2,0);
\draw[dotted, thick] (2,1) --(2,3);
\draw[dotted, thick] (2,1) --(0,0);
\draw[dotted, thick] (2,1) --(4,0);
\filldraw[black] (2,1) circle (1.5pt);
\path (0.6,1.8) node {\tiny $(32)$};
\path (3.4,1.8) node {\tiny $(31)$};
\path (2,-0.4) node {\tiny $(34)$};
\path (2.4,1) node {\tiny $3$};
\path(2,3.3) node {\tiny $D_3$};
\path (-0.3, -0.1) node {\tiny $A_3$};
\path (4.4, -0.1) node {\tiny $B_3$};
\end{tikzpicture}
\end{array}
\quad
\begin{array}{c} 
\begin{tikzpicture}[scale=0.6]
\draw (0,0) -- (2,3) -- (4, 0) --cycle;
\draw[dashed, thick]  (2, 1)-- (1.09, 1.62);
\draw[dashed,thick] (2,1) -- (2.91, 1.62);
\draw[dashed,thick] (2,1)--(2,0);
\draw[dotted, thick] (2,1) --(2,3);
\draw[dotted, thick] (2,1) --(0,0);
\draw[dotted, thick] (2,1) --(4,0);
\filldraw[black] (2,1) circle (1.5pt);
\path (0.6,1.8) node {\tiny $(43)$};
\path (3.4,1.8) node {\tiny $(42)$};
\path (2,-0.4) node {\tiny $(41)$};
\path (2.4,1) node {\tiny $4$};
\path(2,3.3) node {\tiny $A_4$};
\path (-0.3, -0.1) node {\tiny $B_4$};
\path (4.4, -0.1) node {\tiny $C_4$};
\end{tikzpicture}
\end{array}
\]
There is one $C$ tensor in the interaction for each of the four vertices $A, B, C, D$ of the tetrahedron. 
Recall that in our notations,  the vertex $A$ (resp. $B,C,D$)  is opposite to the triangle $1$ (resp. $2,3,4$), so it  lies at the intersection of the three triangles $2, 3, 4$. Each of the fields $\vphi_i$, $i=2,3,4$ contributes to the interactions polynomial with a vertex index $A_i$; 
the resulting three indices $A_2, A_3, A_4$ are then contracted via the coefficients $C_{A_{2}A_{3}A_{4}}$.

\subsection{Algebraic formulation}

Remarkably, this generalization of the GFT model has a very natural algebraic interpretation. 
The structure described above can be understood equivalently 
by  demanding that the extra vertex labels  belong to 
 a {\sl semi-simple $\ast$-algebra}: namely, a unital algebra  $(\cA,\dagger,\Tr_\cA)$, equipped with an 
involution $\dagger$ which is also an algebra anti-homomorphism, together with 
a non-degenerate trace $\Tr_\cA$. 
More details about the structure and the notations are presented in Appendix.

In the following we assume the algebra $\cA$ to be finite dimensional and we denote its dimension by $N$. 
We introduce a basis $\{e_A\}$ labelled by $A \in I$, where $|I| = N$, and denote by 1 the unit element.
We can then define the tensors $h_{AB}$ and $C_{ABC}$ as follows  (see Appendix):
\be \label{algebraic}
h_{AB} =  \Tr_{\cA} (e^\dagger_A e_B),\qquad C_{ABC} =   \Tr_{\cA}( ( e_A  e_B) e_C).
\ee
Whenever the algebra is chosen to be associative, the cyclicity of the trace makes the coefficients 
$C_{ABC} =   \Tr_{\cA}(e_A  e_B e_C)$ cyclically symmetric $C_{ABC} = C_{BCA} = C_{CAB}$.

The idea is to view the extended GFT fields as taking valued in the tensor product $\cA^{\otimes 4}$ of four copies of the algebra, by setting:
\bea
&& \bphi_1(g_{14}, g_{13}, g_{12} ) = \vphi_1^{B C D }(g_{14}, g_{13}, g_{12}) 
(1 \otimes  e_{B}  \otimes  e_{ C } \otimes  e_{ D } ) \crcr
&&\bphi_2(g_{21}, g_{24}, g_{23} ) = \vphi_2^{A C D }(g_{21}, g_{24}, g_{23}) 
(  e_{A} \otimes 1 \otimes  e_{ C } \otimes  e_{ D } ) \crcr
&&\bphi_3(g_{32}, g_{31}, g_{34} ) = \vphi_3^{AB D }(g_{32}, g_{31}, g_{34} ) 
(  e_{A}  \otimes  e_{ B } \otimes 1  \otimes  e_{ D } ) \crcr
&&\bphi_4(g_{43}, g_{42}, g_{41} ) = \vphi_4^{A B C }(g_{43}, g_{42}, g_{41} ) 
( e_{A}  \otimes  e_{ B } \otimes  e_{ C } \otimes 1)  \; .
\eea 
where repeated indices $A, ...$ are implicitly summed over.  We also define the hermitian conjugated fields:
\bea
\bphi^{\dagger}_1(g_{14}, g_{13}, g_{12} ) = \bar\vphi_1^{\bar B\bar  C\bar  D }(g_{14}, g_{13}, g_{12}) 
(1 \otimes  e^{\dagger}_{\bar B}  \otimes  e^{\dagger}_{\bar  C } \otimes  e^{\dagger}_{ \bar D } ) \; ,
\eea 
where $\bar \vphi_i$ denotes the complex conjugate.  
The shift symmetry (\ref{gauge2}) simply reads:
\be
\bphi_{i}(g_{1}g,g_{2}g, g_{3}g) = 
\bphi_{i}(g_{1},g_{2}, g_{3}) \; . \label{gauge3}
\ee
Using this notation, the kinetic term (\ref{kindices}) takes the form:
\be\label{kalg}
K= \sum_{i=1}^4 \int dg_1 dg_2 dg_3 \;\;  |I|^{-1}\, \Tr_{\cA^{\otimes 4}}\Bigl[ \bphi^{\dagger}_i(g_{1}, g_{2}, g_{3} ) \bphi_i(g_{1}, g_{2}, g_{3} )  \Bigr],
\ee
where the factor $|I|^{-1} = \dim(\cA)^{-1}$ is included to compensate the term 
${\rm Tr}_\cA(1)$ showing up in the evaluation of the trace on $\cA^{\otimes 4}$.
The interaction (\ref{intdices}) reads:
\be\label{Valg}
V=  \int \prod_{i<j} \rd g_{ij} 
  \mathrm{Tr}_{\cA^{\otimes 4}}\left[[[\bphi_{1}(g_{14}, g_{13}, g_{12})
\bphi_{2}(g_{21}, g_{24}, g_{23})]
\bphi_{3}(g_{32}, g_{31}, g_{34})]\bphi_{4}(g_{43}, g_{42}, g_{41})\right]. 
\ee
It is straightforward to check that the evaluation of this action reproduces the previous expressions (\ref{kindices},\ref{intdices}).

\subsection{Symmetry}

New symmetries arise from the extension of the GFT model when the algebra $\cA$
characterized by the couplings  is associative. For instance the translation of the vertex $A=(234)$ can be expressed in terms  of three 
unitary operators $\bm U_{23}=\bm U_{32}^{-1}, \bm U_{34}= \bm U_{43}^{-1}, \bm U_{41}=\bm U_{14}^{-1}$ in $\cA$ (unitarity means that $ \bm U^{\dagger}=\bm U^{-1}$).
The non-abelian translational symmetry then reads:  
\bea \label{sym}
\tilde{T}_{A}(\bphi_{1})(g_{14}, g_{13}, g_{12})&=& \bphi_{1}(g_{14}, g_{13}, g_{12}), \nonumber \\
\tilde{T}_{A} (\bphi_{2})(g_{21}, g_{24}, g_{23}) &=& 
\bm U^{1}_{24} \bphi_{2}(g_{21}, g_{24}, g_{23}) \bm U^{1}_{23}, \nonumber\\
\tilde{T}_{A} (\bphi_{3})(g_{32}, g_{31}, g_{34})&=& 
\bm U^{1}_{32} \bphi_{3}(g_{32}, g_{31}, g_{34})\bm U^{1}_{34} ,\nonumber\\
\tilde{T}_{A} (\bphi_{4})(g_{43}, g_{42}, g_{41}) &=&\bm U^{1}_{43} \bphi_{4} (g_{43}, g_{42}, g_{41}) \bm U^{1}_{42}.
\eea
where the superscripts indicate the tensor factor which the unitary elements act on: 
$\bm U^{1} = U^A e_A \otimes 1 \otimes 1 \otimes 1$, $\bm U^{2} = U^B 1 \otimes e_B \otimes 1 \otimes 1$, and so on. 
It can be checked that the action (\ref{kalg},\ref{Valg}) is invariant under this transformation and its analogs for the three other vertices.
This is the non-abelian generalization of the symmetry transformation (\ref{transab2}).

It is also interesting to try to write the non-abelian analog of (\ref{transab}).
To do so, we need the involution algebra $\cA$ to carry a representation of the group $G$. This requires the existence of maps 
$\bm D: G\to \cA$ with $\bm D(g)= D(g)^{A} e_{A}$ that are compatible with the group and unitary structure:
\be
 \bm D(g_1) \bm D(g_2) = \bm D (g_1g_2),\qquad \bm D(g)^{\dagger}= \bm D(g^{\dagger}).
\ee
A  key example of such a structure is when the algebra 
$\cA$ is isomorphic to the algebra ${\rm End}(V)$  of endomorphisms of a  finite dimensional unitary representation $V$ of the group.
In this case $N=d^{2}$ where $d$ is the dimension of the group representation.
We can now write  an analogue of (\ref{transab}), in which the maps  $ \bm D(g) =D^{A}(g) e_{A}$ 
 play the role of non-abelian characters:
\bea
T_{A}(\bphi_{1})(g_{14}, g_{13}, g_{12})&=& \bphi_{1}(g_{14}, g_{13}, g_{12}), \crcr
T_{A} (\bphi_{2})(g_{21}, g_{24}, g_{23}) &=& 
\bm D^{1}({g}_{24}^{-1}) \bphi_{2}(g_{21}, g_{24}, g_{23}) \bm D^{1}({g}_{23}) , \crcr
T_{A} (\bphi_{3})(g_{32}, g_{31}, g_{34})&=& 
\bm D^{1}({g}_{32}^{-1}) \bphi_{3}(g_{32}, g_{31}, g_{34})\bm D^{1}({g}_{34}),\crcr
T_{A} (\bphi_{4})(g_{43}, g_{42}, g_{41}) &=&\bm D^{1}({g}_{43}^{-1})\bphi_{4} (g_{43}, g_{42}, g_{41}) \bm D^{1}({g}_{41}) \; .
\eea
This transformation is a clearly a symmetry of the action.
Note however that it does not preserve the shift invariance (\ref{gauge3}). Instead of being invariant, the image field is now covariant under shift: 
\be
T_{A}(\bphi_{\ell})(g_{1}g,g_{2}g, g_{3}g) = 
\bm D^{1}(g^{-1})T_{A}(\bphi_{\ell})(g_{1},g_{2}, g_{3}) \bm D^{1}(g).
\ee

\section{Feynman amplitudes}
In this section we examine the Feynman expansion of our model. As we will see,  compared to the original colored Boulatov model, the Feynman amplitudes contain an extra contribution from each bubble of the graph. 
In the case where the algebra $\cA$ is associative,  this contribution is a topological invariant characterizing the topology of the bubble. 

\subsection{Colored graphs}

\begin{figure}[ht]
\centering
\includegraphics[scale=1.2]{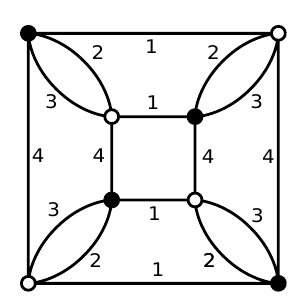}
\caption{Example of a colored graph}\label{fig:colorgr}
\end{figure}

The graphs generated by  the Feynman  expansion of colored group field theories are bipartite edge-colored four-valent graphs \cite{color1}; see Figure \ref{fig:colorgr} for an example. 

\noindent Such a graph $\cG$ possesses a full 3 dimensional cellular structure, where $j$-dimensional cells for $j=0,\cdots,  3$ are defined as the maximally connected $j$-colored subgraphs. 
By (Poincar\'e) duality it also represents a three-dimensional  simplicial complex $\Delta_{\cG}$. Upon this duality, the $j$-cells of the graph correspond to simplices of co-dimension $j$ in 
$\Delta_{\cG}$:  the 0-cells, or vertices of the graph,  correspond to tetrahedra; the 1-cells, or edges of the graph, correspond to triangles; the 2-cells,
or faces of the graph, correspond to edges; and the 3-cells, or `bubbles' of the graph, correspond to the vertices of $\Delta_{\cG}$.

What is remarkable is that the simplicial complexes dual to bipartite edge colored graphs are orientable, triangulated 
normal simplicial pseudo-manifolds \cite{Pezzana}.  `Normal' means that the  link of each simplex of codimension $\geq 2$ (each edge and each vertex in dimension three) is itself a pseudo manifold (hence in particular connected): for example the pinched torus in two dimensions is not a normal pseudo-manifold, because the link of the pinching point is made of two circles. Note that in our case, the link of a simplex of co-dimension $j$ is the pseudo-manifold dual to the corresponding $j$-cell of the graph (which is itself a $j$-colored graph). 

Given a three-dimensional orientable normal pseudo-manifold, the link of each vertex is thus a connected surface of genus $g$.
It is straightforward to check that the sum of genera of the links of all the vertices in a normal simplicial pseudo manifold 
equals the Euler characteristic $\chi = \sum_{i=0}^3 (-1)^i V_i$ (where $V_i$ is the number of simplices of dimension $i$)
of the pseudo manifold. In particular a pseudo-manifold is a manifold if and only if the links of all the vertices are spheres. By duality, this means that a colored graph $\cG$ represents a three-dimensional manifold if and only if its bubbles are all dual to spheres.

\subsection{Graph evaluation}

The interaction vertex $V$ (resp. $\bar{V}$) is represented by a black (white) and positive clockwise turning (negative anti-clockwise turning) vertex:%
\begin{figure}[ht]
\vspace{-0.2cm}
\hspace{1cm}
\begin{minipage}[t]{0.2\textwidth}
\includegraphics[scale=0.1]{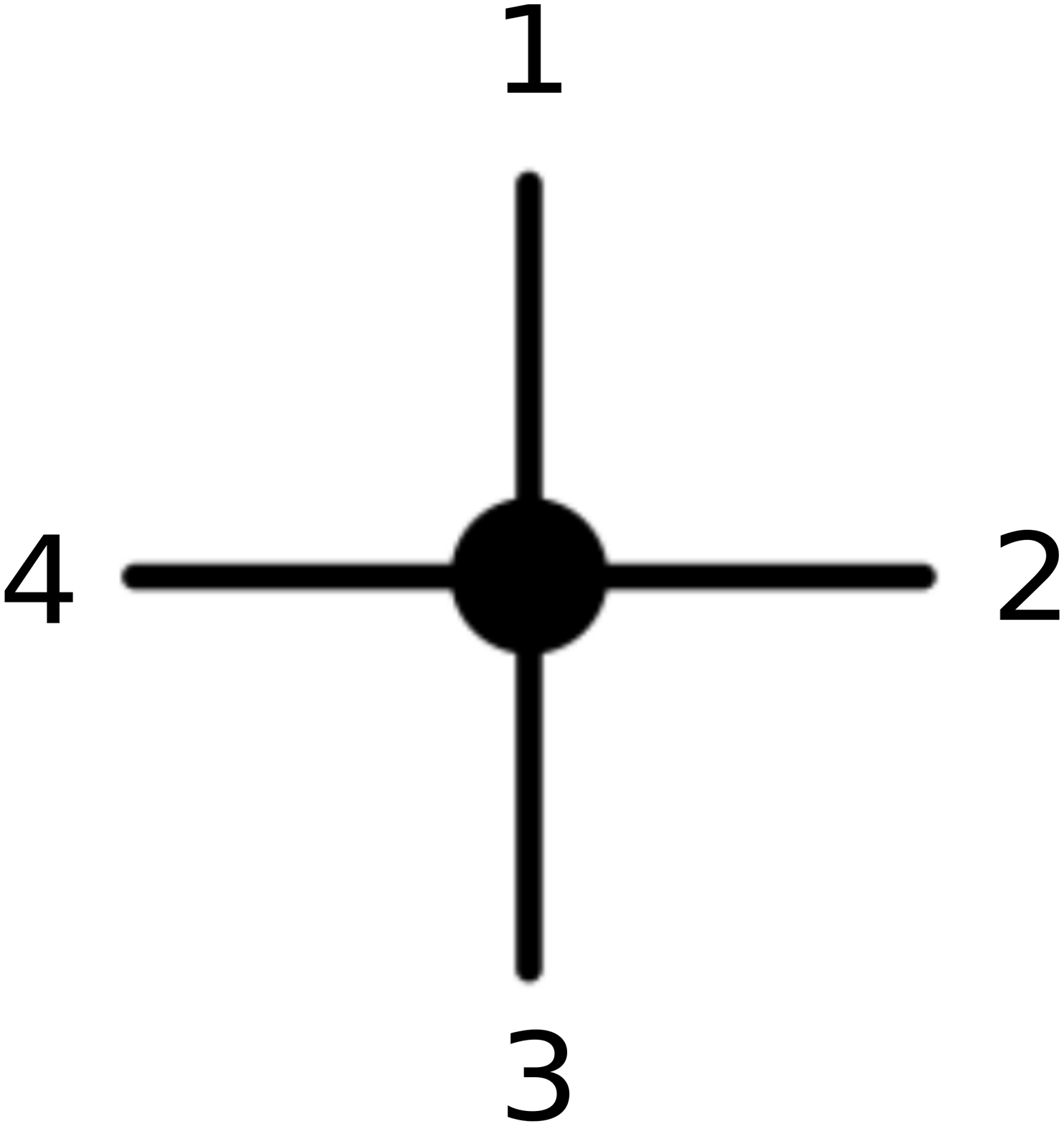}
\end{minipage}
\quad
\begin{minipage}[t]{0.2\textwidth}
\vspace{-2.65cm}
\includegraphics[scale=0.1]{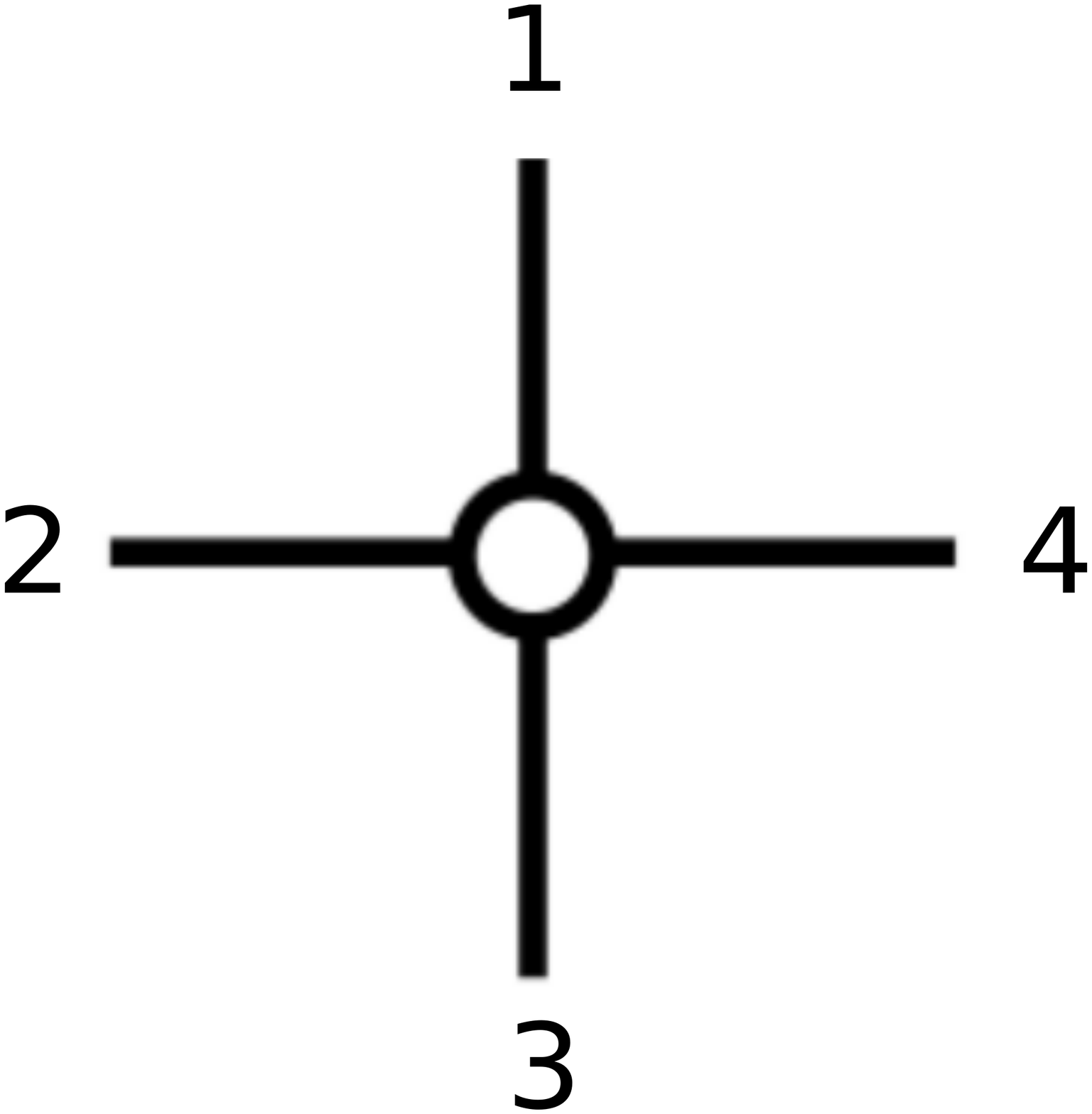}
\end{minipage}
\vspace{-0.5cm}
\centering
\end{figure} 

\noindent Each half-edge of color $i$ incident to a black vertex represents a field $\bphi_i$; it carries three group elements $g_{ij}, g_{ik}, g_{il}$ and three capital letter indices ($B_1, C_1, D_1$ for $i=1$; $A_2$, $C_2$, $D_2$ for $i=2$; and so on).  The  kernel for the vertex $V$ is:
\bea 
V: \quad C_{A_2 A_3 A_4} \, C_{B_1  B_3  B_4}\, C_{C_1 C_2  C_4}\,
  C_{D_1 D_2 D_3} \, \prod_{i<j} \delta\bigl( g_{ij} g^{-1}_{ji}\bigr) \; ,
\eea 
where the interaction coefficients are defined in terms of the trace 
$C_{ABC}:=\Tr_{\cA}(e_{A}e_{B}e_{C})$. A half-edge of color $i$ incident to a white vertex represents a field $\bphi_i^\dag$; it carries three group elements $g_{ij}, g_{ik}, g_{il}$ and three capital letter indices ($\bar{B}_1, \bar{C}_1, \bar{D}_1$ for $i=1$; $\bar{A}_2$, $\bar{C}_2$, $\bar{D}_2$ for $i=2$; and so on). The kernel for the dual vertex $\bar V$ is
\be
\bar V: \quad \wc_{\bar A_4 \bar A_3 \bar A_2} \, \wc_{\bar B_4  \bar B_3  \bar B_1}\, \wc_{\bar C_4 \bar C_2  \bar C_1}\,
  \wc_{\bar D_3 \bar D_2 \bar D_1} \, \prod_{i<j} \delta\bigl( g_{ij} g^{-1}_{ji}\bigr). \; 
\ee
where the coefficients are defined in terms of the trace of conjugate algebra elements 
$\wc_{ABC}\equiv \Tr(e_{A}^{\dagger} e_{B}^{\dagger} e_{C}^{\dagger}) $
Due to the anti-homomorphism property of the involution, the two sets of coefficients are related by complex conjugation and odd index permutation, i.e $\wc_{ABC} =  \overline{C_{CBA}}$.

Taking into account the invariance of the field Equ. \eqref{gauge2}, each colored edge of the graph (oriented from $\bar{V}$ to $V$)  carries the propagator:
\bea 
G: \quad {h}^{ A \bar A}\, {h}^{ B \bar B}\, {h}^{ C \bar C} \int dh \; \delta^{\Lambda} \bigl(g_{1} h (g'_{1})^{-1}\bigl)
\delta^{\Lambda} \bigl(g_{2} h (g'_{2})^{-1} \bigl) \delta^{\Lambda} \bigl(g_{3} h (g'_{3})^{-1} \bigl) \; ,
\eea
where $ \delta^{\Lambda}$ denotes a cut-off delta function\footnote{chosen to be a class function $\delta^{\Lambda}(k gk^{-1}) = \delta^\Lambda(g)$.} 
on the group and ${h}^{A\bar A}$ is the inverse of the hermitian form (\ref{algebraic}), 
\be \label{hinverse}
{h}^{A\bar A} h_{\bar A A'} = \delta_{A'}^A.
\ee

Given a closed connected Feynman graph ${\cal G}$, we denote by ${\cal V}_{{\cal G}}$, ${\cal E}_{{\cal G}}$, ${\cal F}_{{\cal G}}$ and ${\cal B}_{\cal G}$
be the sets of its vertices, edges, faces and bubbles. 
Our first result is that the contribution due to the new  algebraic structure we have introduced, factorizes as a product over the bubbles $b\in\cB_{\cal G}$ 
of the graph. In fact,  the bubbles being defined as the maximally connected 3-colored subgraphs, 
each vertex of $\cG$ belongs to exactly four bubbles of colors $A = (234), \, B = (134), \, C = (124) $ and $D = (123)$ and therefore contributes a factor:
\be
C_{A_2A_3A_4} \, \cdots \,\, C_{D_{1}D_{2}D_{3}},\quad \mathrm{or} \quad
\wc_{\bar A_4 \bar A_3 \bar A_2} \, \cdots \,\, \wc_{\bar{D}_{3}\bar{D}_{2}\bar{D}_{1}} 
\ee
for a vertex of type $V$ or $\bar{V}$ respectively.  Each edge of $\cG$ belongs to exactly three bubbles, and contributes 
an independent  factor for each of them: for instance an edge of color 1 contributes  $h^{ B \bar B}$ for the bubble  $B = (134)$, $h^{ C \bar C}$ for 
the bubble $C = (124)$, and $h^{ D \bar D}$ for the bubble $D = (123)$. Therefore the edge contributes the total factor:
  \be
  h^{ B \bar B}h^{ C \bar C}h^{ D \bar D} \; .
  \ee
to the amplitude. The evaluation of $\cG$ is thus the product of two factorized contributions:
\be \label{eval}
A(\cG, \Lambda) = \lambda^{2|\cal V_{\cal G}|}A_0(\cG, \Lambda) \, \left(\prod_{b\in {\cal B}_{\cG}} a(b)\right) \; ,
\ee
where $|\cal V_{\cG}|$ counts the number of vertices of the graph and 
the amplitude $A_0(\cG, \Lambda)$ is the evaluation of the graph $\cG$ in the non-modified Boulatov theory:
\bea \label{boul}
 A_0(\cG;\Lambda) =  \int  \Bigl(\prod_{e\in \cE_{\cal G}} dh_e\Bigr) \prod_{f\in \cF_{\cal G}} \delta^{\Lambda} \Bigl( \prod^{\rightarrow}_{e\in \partial f}
 h_e^{s(e,f)} \Bigr) \; .
\eea 
To obtain this expression for $A_0$, all edges of $\cG$ have been oriented from the vertex $\bar V$ to the vertex $V$, and each face has been assigned an arbitrary orientation. 
The integrals are over all assignments of a group element $h_e \in G$ to each edge $e$ of the graph.  
The arguments of the cut-off delta functions are ordered products of group elements along the boundary $\partial f$ of the faces $f$; $s(e, f)$ is the 
edge-face adjacency
matrix of the graph: $s(e, f)= 1$ or $-1$ 
depending on whether the orientations of $e$ and $f$ match or not\footnote{If one interprets the assignment $\{h_e\!\! \in\!\! G\}_{\e\in \cG}$ as a discrete connection
on the graph, 
the integrand in (\ref{boul}) imposes, in the limit of large cut-off, that the holonomy along the loop bounding each face is trivial.  The Boulatov 
amplitude formally gives an integral over the space ${\rm Hom} (\pi_1(\cG), G)$ of flat discrete connections on $\cG$.}. 

The contribution  $a(b)$ of a bubble $b \in{\cal B}_{\cG}$ of colors $A=(234)$ is obtained by tracing out   along $b $ all the tensors $C,\bar{C}$  and $h$ 
carrying $A$-indices:
\bea \label{bub}
 a(b) = \sum_{\{A, \bar{A}\}}  \prod_{(v, \bar v)  \in  {\cal V}_b} C_{A_2^v A_3^v A_4^v} \, \wc_{\bar A_4^{\bar v}\bar A_3^{\bar v}\bar A_2^{\bar v}}
  \prod_{e_i \in  {\cal E}_b } 
  {h}^{ A_i^{v}  \bar A_i^{\bar v} } \;,
\eea 
The products run over the vertices $v, \bar{v}$ and oriented edges $e_i =(\bar v , v)$ of the bubble $b$; and the sum runs over a set of indices 
$A_i^v,  \bar A_i^{\bar v}$ labeled by a vertex  of the bubble and a color $i=2,3,4$. 
We recognize here the trace invariant\footnote{This is an invariant under the action of $U(N)^3$ corresponding to independent unitary
(with respect to the form $h$) 
transformations on each index $A$.} \cite{universality} associated to the bubble $b$ and built out of the tensor $C$.

\subsection{Bubble weights and topological invariants}

The goal of this section is to point out that, when the algebra $\cA$ is chosen to be associative, the weight (\ref{bub}) defines a topological invariant of
the surface represented by the bubble $b$. In fact, as mentioned above, a bubble (or 3-cell) $b$ of the graph  is itself a 3-colored graph dual to a triangulated Riemann surface $\Sigma_b$. Specifically, in terms of the dual pseudo-manifold $\Delta_{\cG}$, this surface is the {\sl link} of the vertex $v \in \Delta_{\cG}$ dual to $b$. It can be visualized as the surface triangulated by the triangles
opposite to $v$ in all tetrahedra  $\Delta_{\cG}$ that have $v$ as a vertex. By duality the vertices of a bubble $b$ correspond to the triangles of $\Sigma_b$, and the edges of $b$ correspond to the edges of $\Sigma_b$.  
Note that the triangulation of $\Sigma_b$ inherits a bipartite set of triangles and a colored set of edges. Moreover the surface is oriented: the colors give 
a consistent ordering $(e_1, e_2, e_3)$ of the boundary edges of each triangle. 

Using this correspondence, our observation is that the weight $a(b)$  takes the form of the partition function of a lattice model on $\Sigma_b$, with the tensors
$C_{ABC}, \wc_{\bar A \bar B\bar C}$ as 3-point vertices and the (inverse) hermitian form $h^{A\bar B}$ as propagator.  We have in fact that
$ a(b) = a(\Sigma_{b})$, where the amplitude $a(\Sigma)$ for a (bipartite, edge colored) triangulated surface $\Sigma$ is given by
\be \label{lattice}
a(\Sigma) = \sum_{\{A_e^t\}} \, \prod_{t, \bar{t}: \,  \Sigma-\mbox{\footnotesize triangles}} C_{A_{e_1}^t A_{e_2}^t A_{e_3}^t} 
\wc_{\bar{A}_{e_3}^{\bar{t}} \bar{A}_{e_2}^{\bar{t}} \bar{A}_{e_1}^{\bar{t}}} 
\prod_{e: \, \Sigma-\mbox{\footnotesize edges}}  {h}^{A_e^{t_1} \bar A_e^{\bar t_2}} 
\ee
The sum is over a set of indices  $A_e^{t}$ labeled by a pair $(t, e)$, where $t$ is a triangle of $\Sigma$ and $e\subset \partial t$ is an edge of $t$. 

Note that for  a triangulated surface $\Sigma$, the number of edges $|E|$ and the number positively (resp. negatively) oriented vertices
$|V|, |\bar{V}|$ are related as $2 |E| = 3|V| + 3|\bar{V}|$. This shows that the evaluation $a(\Sigma)$ is invariant under the rescaling
\be
C_{ABC} \to \alpha^{3} C_{ABC},\qquad \wc_{\bar{C}\bar{B} \bar{A}}\to \bar{\alpha}^{3}\wc_{\bar{C}\bar{B} \bar{A}},\qquad h_{\bar A A}\to |\alpha|^{2} h_{\bar A A}.
\ee
This transformation can be reabsorbed in  an algebra basis redefinition $e_{A}\to \alpha e_{A}$, which corresponds to a field rescaling.

We have written the amplitude in terms of the inverse of the hermitian form $h_{\bar{A}A}= \Tr( e^{\dagger}_{\bar{A}} e_{A})$ and the couplings $C,\wc$. 
It is also convenient to write the same amplitude in terms of the metric
\be
g_{AB} \equiv \Tr(e_{A} e_{B}),
\ee
and the coupling $C$ only. Let us denote by  $ g^{AB}$ the inverse metric.
As recalled in Appendix, the action of the involution on the basis elements reads:
\be
 e^{\dagger}_{\bar{A}} = h_{\bar{A} A } g^{A A'} e_{A'}.
\ee
Now since  the interaction coefficients are given by
\be
C_{ABC} = \Tr_{\cA}(e_A e_B e_C), \qquad   \wc_{\bar A \bar B\bar C} = \Tr_{\cA}(e_{\bar A}^{\dagger} e_{\bar B}^{\dagger} e_{\bar C}^{\dagger}), 
\ee
the following relation holds: 
\be \label{C-rel}
\wc_{\bar A \bar B\bar C}  = h_{\bar{A} A}h_{{\bar B B}} h_{\bar{C} C}g^{A A'}g^{B B'}g^{C C'}\, C_{A'B'C'} .
\ee
By plugging this relation into the amplitude (\ref{lattice}) and using the inverse relation (\ref{hinverse}), we obtain:
\be \label{lattice-g}
a(\Sigma) = \sum_{\{A_e^t\}} \, \prod_{t: \, \Sigma-\mbox{\footnotesize triangles}} C_{A_{e_1}^t A_{e_2}^t A_{e_3}^t}\, 
\prod_{e: \, \Sigma-\mbox{\footnotesize edges}}  g^{A_e^{t_1} A_e^{t_2}}
\ee
where the product of $C$ tensors runs over all oriented triangles of $\Sigma$; the ordering of indices in each $C_{A_{e_1}^t A_{e_2}^t A_{e_3}^t}$ 
is determined by  the orientation of the triangle $t$.

\subsubsection{Topological invariance}

Written in the form (\ref{lattice-g}),  the tensor invariant reproduces the Fukuma, Hosono, Kawai definition of a 2d (lattice) topological
field theory \cite{FHK,Turaev}. In fact, assuming our semi-simple algebra $\cA$ is also associative, the weight (\ref{lattice-g}) defines a 
topological invariant of the surface $\Sigma$. 
This is because the conditions of associativity and semi-simplicity (i.e invertibility of the metric $g_{AB}$) are precisely the conditions 
which guarantee that the amplitude $a(\Sigma)$ is independent of the triangulation of $\Sigma$. Indeed as recalled in Appendix, under such conditions
the metric $g_{AB}$ and coefficients $C_{ABC}$ can be written in terms of the algebra structure constants $e_{A}e_{B}=C_{AB}{}^{C} e_{C}$ as: 
\be \label{C-g}
  g_{AB}\equiv C_{A E}{}^{F}C_{B F}{}^{E}, \qquad \quad C_{ABC}= C_{AB}{}^{D}g_{DC}.
\ee
Now, the topological invariance of $a(\Sigma)$ amounts to its invariance under the two local Pachner moves of the triangulation, which translates algebraically as:
\bea
\begin{array}{c}
\begin{tikzpicture}
\draw (0,0) rectangle (1, 1);
\draw (0,0) --(1, 1);
\draw[dashed] (-0.2,0.7) -- (0.25, 0.7);
\draw[dashed] (0.25, 0.7) -- (0.25, 1.2);
\draw[dashed] (1.2,0.3) -- (0.75, 0.3);
\draw[dashed] (0.75, 0.3) -- (0.75, -0.2);
\draw[dashed] (0.25, 0.7)--(0.75, 0.3);
\path (-0.35, 0.7) node {\tiny $A$};
\path (0.25, 1.35) node {\tiny $B$};
\path (0.75, -0.35) node {\tiny $D$};
\path (1.35,0.3) node {\tiny $C$};
\end{tikzpicture}
\end{array} &=&   \begin{array}{c}
\begin{tikzpicture}
\draw (0,0) rectangle (1, 1);
\draw (1,0) -- (0, 1);
\draw[dashed] (1.2,0.7) -- (0.75, 0.7);
\draw[dashed] (0.75, 0.7) -- (0.75, 1.2);
\draw[dashed] (-0.2,0.3) -- (0.25, 0.3);
\draw[dashed] (0.25, 0.3) -- (0.25, -0.2);
\draw[dashed] (0.75, 0.7)--(0.25, 0.3);
\path (-0.35, 0.3) node {\tiny $A$};
\path (0.75, 1.35) node {\tiny $B$};
\path (0.25, -0.35) node {\tiny $D$};
\path (1.35,0.7) node {\tiny $C$};
\end{tikzpicture}
\end{array}
\qquad C_{AB}{}^{E}C_{EC}{}^{D} = C_{BC}{}^{F} C_{AF}{}^D  \; , \label{2-2}
\\
\begin{array}{c}
\begin{tikzpicture}
\draw (0,0) -- (0.75,1) -- (1.5, 0) --cycle;
\draw[dashed] (0.75, 0.35) -- (0.75, -0.2);
\draw[dashed] (0.75, 0.35) -- (0.3, 0.7);
\draw[dashed] (0.75, 0.35) -- (1.2, 0.7);
\path (0.2, 0.8) node {\tiny $A$};
\path (1.3, 0.8) node {\tiny $B$};
\path (0.75, -0.35) node {\tiny $C$};
\end{tikzpicture}
\end{array}
&=&
\begin{array}{c}
\begin{tikzpicture}
\draw (0,0) -- (0.75,1) -- (1.5, 0) --cycle;
\draw (0.75, 0.35) -- (0.75,1);
\draw (0.75, 0.35) --(0,0);
\draw (0.75, 0.35) -- (1.5, 0);
\draw[dashed] (0.75, 0.15) -- (0.75, -0.2);
\draw[dashed] (0.55, 0.5) -- (0.3, 0.7);
\draw[dashed] (0.95, 0.5) -- (1.2, 0.7);
\draw[dashed] (0.55, 0.5) -- (0.95, 0.5);
\draw[dashed] (0.55, 0.5) -- (0.75, 0.15);
\draw[dashed] (0.95, 0.5) -- (0.75, 0.15);
\path (0.2, 0.8) node {\tiny $A$};
\path (1.3, 0.8) node {\tiny $B$};
\path (0.75, -0.35) node {\tiny $C$};
\end{tikzpicture}
\end{array}
\qquad \qquad C_{ABC} \, = \,   C_{AD}{}^E C_{BF}{}^D C_{EC}{}^F \; .
\eea
It is straightforward to check that the first equation expresses the associativity of $\cA$ and the second one follows from the relations (\ref{C-g}).

Since any semi-simple associative algebra is a direct sum of matrix algebras, 
the basic example for us is the algebra $\cA =\mbox{End(V)} = \mbox{Mat}_{d\times d}(\mathbb{C})$ of endomorphisms of a finite dimensional vector space $V$. 
In this case, the bubble weight $a(b)$ is just the evaluation of the Feynman amplitude of the graph dual to the surface $\Sigma_b$ 
of a cubic matrix model, and is given by:
\be
a(b) = d^{\chi(\Sigma_{b})} \; ,
\ee
where $d$ is the dimension of $V$ and $\chi(\Sigma_{b}) = 2- 2g_{b}$ is the Euler characteristic of the surface written in terms of its genus $g_b$.  
Plugging this value into the Feynman evaluation (\ref{eval}) of the graph $\cG$ gives:
\be\label{amp1}
A(\cG;\Lambda) = {\lambda}^{ 2|{\cal V}_{\cG}|} d^{2|B_{\cG}|}A_0(\cG;\Lambda) \, \prod_{b\in \cB_{\cG}}d^{-2 g_{b}}.
\ee

Another classical example of associative algebra is the group algebra $\cA=\mathbb{C}[H]$ of a finite group $H$ with cardinal $N=|H|$.  The set
of indices is $H$ and we choose:
\be
C_{AB}{^C} = N  \, \delta(AB, C), \qquad g_{AB} = N\, \delta(AB,1) \;,
\ee
where $\delta$ is the Kr\"onecker delta on $H$. In this case one can show:
\be\label{aG}
a(b) = N^{\chi(\Sigma_b) -1} |\mbox{Hom}(\pi_1(\Sigma_b), H)|.
\ee
where $\pi_1(\Sigma_b)$ denotes the first homotopy group of the surface $\Sigma_b$. Note that we can also represent the algebra $\mathbb{C}[H]$ 
in terms of a sum over its irreducible representations:
\be
\mathbb{C}[H] \simeq \bigoplus_{V\in \hat{H} } \mathrm{End}_{\mathbb{C}}(V)
\ee
where 
$ \hat{H}$, denoted the set of isomorphism classes of irreducible representations of $H$.  This means that we can also evaluate the amplitude as: 
\be\label{aV}
a(b) = \sum_{V \in \hat{H}} \rd_{V}^{\chi(\Sigma_{b})}.
\ee
The equality between this evaluation and (\ref{aG}) is the Mednykh's formula \cite{Synder}. In the case of the sphere,  this formula gives the dimension:
\be
N=\sum_{V \in \hat{H}} \rd_{V}^{2}.
\ee

\subsubsection{Manifolds v.s pseudo-manifolds}

What is remarkable with the amplitudes (\ref{amp1}) is that the new contribution from the bubbles allows us to control the relative weight of manifolds 
versus pseudo-manifolds in the Feynman expansion (see \cite{DanieleSylvain} for a related study of this problem).  
In fact, as we have recalled, the condition for the pseudo-manifold $\Delta_{\cG}$ represented by the graph $\cG$ to be a manifold is that its Euler characteristic
\be \label{Euler3d}
\chi_{\cG} \equiv {-|\cal V_{\cG}|+ |\cal E_{\cG}| - |\cal F_{\cG}| + |\cal B_{\cG}|} = \sum_{b \in \cal B_{\cG}} g_{b} \;,
\ee
vanishes. This amounts to requiring that all bubbles have spherical topology: $g_b = 0, \forall b\in \cB_{\cG}$.  
It is clear from the expression (\ref{amp1}) that,  at fixed number of bubbles,  the amplitudes of pseudo-manifolds are suppressed in the large $d$ limit.

 There is another positive integer characterizing the geometry of the dual pseudo-manifold $\Delta_{\cG}$,  called the {\sl degree}  $\omega_{\cG}$ \cite{Raz2}.
 This integer has a beautiful geometrical interpretation first  given by Ryan in \cite{Ryan} in the case of a manifold. 
 This intuitively goes as follows: as we have seen,  each vertex of $\Delta_{\cG}$ (dual to a bubble of $\cG$) carries a color index
 $A,B,C,D$.
The subset of all vertices of color $A$ or $B$, together with the edges of color $AB$, form a subgraph of the 1-skeleton of 
$\Delta_{\cG}$.  Let $\Sigma_{AB}$ be the surface defined as the boundary of a neighborhood of this subgraph in $\Delta_{\cG}$. 
It can also be  thought of as the surface obtained by gluing pairwise the (holed) surfaces $\Sigma_b$ associated to all bubbles $b $ of
color $A$ or $B$, each cut along attaching disks  
along which we glue cylinders corresponding to edges of color $AB$.  
 By construction the Euler characteristic of $\Sigma_{AB}$ is given by:
\be \label{chiAB}
\chi_{AB} = \sum_{b: \, \mbox{\tiny color $\!\!A$ \!or\! $B$}} \chi(\Sigma_b)  - 2 | {\cal F}_{AB}|,
\ee
where $| {\cal F}_{AB}|$ is the number of edges of $\Delta_{\cG}$ (or equivalently faces of $\cG$) of color $AB$. 
One can also show that $\Sigma_{AB}=\Sigma_{CD}$; in fact this surface splits $\Delta_{\cG}$ in two connected components.  
If one denote by $g_{AB}$ the genus of the surface $\Sigma_{AB}$,
so that $\chi_{AB} = 2-2g_{AB}$, we then have that $g_{AB}=g_{CD}$.
In the manifold case,
$\Sigma_{AB}$  is a Heegard splitting surface for $\Delta_{\cG}$. The {\sl degree} is defined as:
 \be
 \omega_{\cG} \equiv g_{AB}+ g_{AC}+ g_{AD}= \frac12 \sum_{(MN)}g_{MN} \; ,
 \ee
 where the sum is over all possible pairs of colors.  
The relations (\ref{chiAB}) and (\ref{Euler3d}) then lead to the key formula for the degree \cite{Raz2}:
 \be 
|V_{\cG}|-  2|B_{\cG}| + 2\chi_{\cG}   = 2(\omega_{\cG} -3).
 \ee
Thus,  $2(\omega_{\cG}-3)$ measures the difference between the number of vertices and  the sum of the Euler characteristics of all bubbles.  
Therefore, defining a new coupling 
$
 g^{2} \equiv  \lambda^{2} d,
$
we can write the amplitude (\ref{amp1}) as 
 \be\label{amp2}
A(\cG;\Lambda) = {g}^{ 2|{\cal V}_{\cG}|} d^{2(3 - \omega_{\cG})}A_0(\cG;\Lambda).
\ee
We recover the familiar suppression of complexes with high degree in the limit of large $d$ \cite{Raz2}.

\subsubsection{Dipole moves}

We discussed so far the form (\ref{lattice-g}) of the tensor invariant associated to each bubble of the graph.  We recognized the partition 
function of a topological lattice model built from a  semi-simple algebra with structure constant $C$ and metric $g_{AB}$. Now, the same 
invariant can also be written in terms of the {\sl bipartite} structure as in (\ref{lattice}). In this section, we would like to extend the discussion of topological invariance to the case of bipartite triangulations. The Pachner moves are in fact no longer adapted to this case, since they do not preserve the bipartite structure.

\begin{figure}[ht]
\centering
\includegraphics[width=12cm]{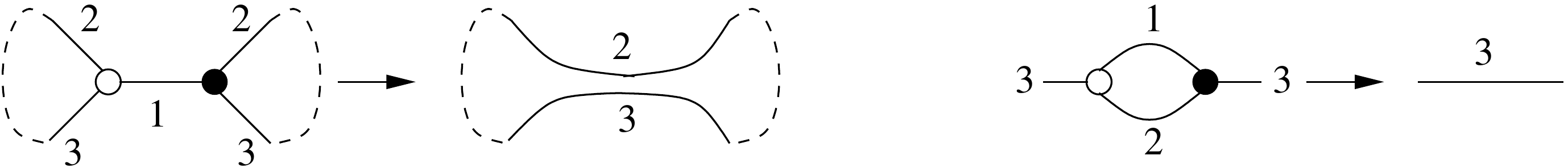}
\caption{Dipole moves}\label{fig:dipole}
\end{figure}

The analogue of the Pachner moves for bipartite triangulations are the so-called dipole moves \cite{dipole}. These give rise to an analogue of
the Pachner theorem: any two bipartite triangulations of the same topological surface are related by a sequence of dipole moves.  
Consider a graph with edges colored $c_1,c_2$ and $c_3$. A $1$-dipole of color $c$ is an edge of color $c$ 
joining two vertices $v$ and $\bar v$ such that $v$ and $\bar v$ belong to two \emph{different} 
faces of colors $\{c_1,c_2,c_3\} \setminus \{c\}$. A $2$-dipole of colors $c,c'$ is a pair of edges of colors 
$c$ and $c'$ joining two vertices $v$ and $\bar v$ (such that the edges of color $\{c_1,c_2,c_3\} \setminus \{c,c'\}$
hooked to $v$ and $\bar v$ are different). As illustrated in Figure \ref{fig:dipole},  a {\sl dipole move} consists of erasing the edge(s) and vertices forming a dipole and reconnecting the remaining edges according to their color. 

To conveniently write down the invariance under these moves, let us introduce the matrices
$C_{A},\wc_{\bar A} \in \mathrm{End}(\cA)$ defined by
\be
(C_{A})_{B}{}^{\bar{C}} \equiv C_{ABC} h^{C\bar C},\qquad
(\wc_{\bar{A}})_{\bar B}{}^{{C}} \equiv \wc_{\bar A\bar C\bar B} h^{C\bar C}.
\ee
These matrices show up in the products $ e_{A}e_{B} = C_{AB}{}^{\bar{C}} e^{\dagger}_{\bar C}$ and $ e^{\dagger}_{\bar A}e^{\dagger}_{\bar B} =
(\wc_{\bar{A}})_{\bar B}{}^{{C}} e_{C}$.
The  invariance under $2$-dipole moves translates into the identity:
\be 
\Tr(\wc_{\bar A} C_{A} ) = h_{ \bar A A}.
\ee
This corresponds to the definition of the hermitian form.
The invariance under $1$-dipole moves is encoded into the relations: 
\bea
& &\Tr\left( \wc_{\bar A_{0}} [C_{A_{1}}\wc_{\bar A_{1}} \cdots C_{A_{m}}\wc_{\bar A_{m}}] C_{B}\right)h^{B \bar B}
\Tr\left( \wc_{\bar B} [C_{A_{m+1}} \wc_{\bar A_{m+1}} \cdots C_{A_{n}} \wc_{A_{n}}] C_{A_{0}}\right)\\
&=& 
\Tr\left( \wc_{\bar A_{0}}[C_{A_{1}}\wc_{\bar A_{1}} \cdots C_{A_{n}} \wc_{A_{n}}] C_{A_{0}}\right).
\eea
Note that the $1$-dipole move is trivially satisfied when the two faces of colors $\{c_1,c_2,c_3\}$ to which $v$ and $\bar v$ belong
consist each of two edges. 
\be
h_{\bar{A}B} h^{B\bar B}h_{\bar B A} = h_{\bar A A}.
\ee

Let us now look at the dipole moves of higher order. The dipole move of length $(n,m)$ is the one containing $n$ $C\wc$ on the left and $m$ $\wc C$ on the right. It is easy to see that the inverse property (\ref{hinverse}) implies all the moves $(n,1)$. To study the higher moves,  it is convenient to notice that the operator $C_{A}\wc_{\bar A}$ represents the left-right multiplication operator:
\be
e_{A} e_{B} e^{\dagger}_{\bar A} = [C_{A}\wc_{\bar A}]_{B}{}^{B'} e_{B'},\qquad
e_{A} e_{\bar B}^{\dagger} e^{\dagger}_{\bar A} = [\wc_{\bar A}C_{A}]_{\bar B}{}^{\bar B'} e_{\bar B'}.
\ee
which implies that 
\bea
[C_{A}\wc_{\bar A}][C_{B}\wc_{\bar B}] &=& C_{BA}{}^{\bar C} \wc_{\bar A \bar B}{}^{C}
[\wc_{\bar C}\wc_{ C}],\crcr
{[}\wc_{\bar A} C_{A}][\wc_{\bar B} C_{B}] &=& C_{BA}{}^{\bar C} \wc_{\bar A \bar B}{}^{ C}
[\wc_{\bar C} C_{C}].
\eea
By using these identities repeatedly,  one can see that the  dipole identity reduces to the $(2,2)$ dipole  move.
Now, if the algebra is semi-simple,  we have the identity
\be \label{dipole h}
\bar{h}^{AB} \Tr_{\cA} (M^\dagger e_B^\dagger ) \Tr_{\cA} (e_A N) = \Tr_{\cA} (M^\dagger N) 
\ee
which implies the dipole moves.
The semi-simplicity condition is therefore a {\it sufficient} condition for implementing the dipole moves.
It is however not clear to us that it is also a necessary condition, i-e whether the invariance under $(2,2)$ dipole  moves requires  semi-simplicity.
If it does not,  that would be suggest the possibility to use bipartite structures to construct new topological models {\sl not} based on semi-simple algebras.   We leave this interesting question for future work. 

%
%
%

\section{Outlook}

We conclude with some possible avenues for future investigation. First, it will be interesting to extend the procedure to higher dimensions: one can assign extra indices to the field, associated to lower dimensional simplices (both vertices and edges in four dimensions). We expect the structure of the translational symmetry to be more involved in higher dimensions. In four dimensions for instance, we expect edge and vertex translations to be intertwined, as the translation of two vertices joined by an edge also translates the edge itself. Studying such higher dimensional generalizations of the model we presented here is left for future work.

Several other generalizations of the construction should be investigated. For instance models based on non-associative algebras will be worth studying. 
As well, the idea of using extra field labels  to assign specific weights to a class subgraphs in the Feynman evaluations may be applied to other subgraphs than the bubbles. Thus an important class  of subgraphs are the so-called jackets, which in the 3d case represent (Heegaard) splitting surfaces for the dual triangulation. Following a similar procedure as above, it will be straightforward to construct a GFT assigning a topological weight to each jacket graph in the Feynman amplitudes. The hope is that a generalization of such constructions can provide us with new tools for a  better control of the topologies generated by the Feynman expansions of group field theories.

 \vspace{0.5cm}

~\\
{\bf \large Acknowledgements}
~\\

We thank Sylvain Carrozza, Maite Dupuis and Florian Girelli for insightful comments and discussions. A.B gratefully acknowledges support of the A.\ von Humboldt Stiftung through a Feodor Lynen Fellowship. Research at Perimeter Institute is supported by the Government of Canada through Industry Canada and by the Province of Ontario through the Ministry of Research and Innovation. This research was also partly supported by grants from NSERC.

\appendix

\section{Glossary on $\ast$-algebras}\label{sec:algebra}

We consider  a finite dimensional algebra $\cA$ over the complex numbers $\mathbb{C}$, equipped with an anti linear involution 
$\cA \to \cA$, $M  \mapsto M^\dagger$ acting as complex conjugation $(MN)^\dagger = N^\dagger M^\dagger$. Given  $M \in \cA$,  
we denote by $\Tr_{\cA}(M) \in \mathbb{C}$ the trace of the multiplication map  $\mathrm{ad}_M \maps \cA \to \cA$ sending any $N$ to $MN$.  
This provides $\cA$ with a two bilinear forms $\cA \otimes \cA \to \mathbb{C}$ defined by 
\be \label{bil}
g(M, M') = \mathrm{Tr}_{\cA}(M M') \qquad h(M, M') = \mathrm{Tr}_{\cA}(M^{\dagger} M'),
\ee
In a given basis $\{e_{A}\}$  of $\cA$ with associated structure constants 
$e_A e_B = C_{AB}{}^{C} e_C$,  
the forms (\ref{bil}) define a symmetric tensor and a hermitian tensor: 
\be \label{tensors}
 g_{AB} := g(e_A, e_B) = \Tr_{\cA} (e_A e_B), \qquad h_{AB} := h(e_A, e_B) = \Tr_{\cA} (e^\dagger_A e_B)
\ee
Since the multiplication map $\mathrm{ad}_{M} \maps \cA \to \cA$ with $M=e_Ae_B$ acts on basis elements as:
\be
\mathrm{ad}_{e_Ae_B} e_C = C_{AB}{}^{E} C_{EC}{}^D e_D,
\ee
its trace, and hence $g_{AB}$, can be written as a  contraction of the structure constants:
\be \label{g-C}
g_{AB}=  C_{AB}{}^{E}C_{ED}{}^{D} 
\ee
Note that this relation fixes the relative normalization of the metric and structure constants.
 We also introduce the conjugation isomorphism:
\be \label{conj}
e_{\bar A}^{\dagger} = h_{\bar A}{}^{B} e_{B}, \qquad 
e_{A} = \bar{h}_{A}{}^{\bar B} e_{\bar B}^{\dagger},\quad
\bar{h}_{A}{}^{\bar B} h_{\bar B}{}^{C} = \delta_{A}^{C}.
\ee
where $\bar{h}_A{}^{\bar B}$ denotes the complex conjugate $\overline{h_{ A}{}^{\bar B}}$. This conjugation matrix can be written in terms of the metrics $h$ and $g$,  in particular:  
\be
h_{\bar AB} = h_{\bar A}{}^{A'} g_{A'B}, \qquad g_{AB} = \bar{h}_{A}{}^{\bar A} h_{\bar AB}
\ee
Upon conjugation, the structure constants $e_{A}^{\dagger}e_{B}^{\dagger} = \wc_{AB}{}^{C}e_{C}^{\dagger}$ are expressed in terms of 
$C_{AB}{}^{C}$ as:
\be
\wc_{AB}{}^{C} \equiv h_{A}{}^{A'}h_{B}{}^{B'}\bar{h}_{C'}{}^{C} C_{A'B'}{}^{C'} = \overline{{C}_{BA}{}^{C}}.
\ee

\subsection{Associativity}

Assuming that $\cA$ is associative, the associativity property translates into the following relations for the structure constants:  
\be \label{assoc}
C_{AB}{}^{E}C_{EC}{}^{D} = C_{BC}{}^{E'} C_{AE'}{}^D  
\ee
Together with (\ref{g-C}), associativity leads to the standard form for the ``Killing'' metric of the algebra: 
\be \label{killing}
g_{AB}=  C_{AC}{}^{D}C_{BD}{}^{C} 
\ee
Associativity also says that the set of coefficients
\be
C_{ABC}\equiv C_{AB}{}^{C'} g_{C'C} =   \Tr_{\cA}(e_Ae_Be_C) 
\ee
are cyclically  symmetric   $C_{ABC} = C_{BCA} = C_{CAB}$. Upon conjugation, we also get:
\be
\wc_{{A}{B}{C}} \equiv \wc_{AB}{}^{C} g_{C'C} = \Tr_{\cA}(e_{A}^{\dagger}e_{B}^{\dagger}e_{C}^{\dagger})=  \overline{{C}_{CBA}}.
\ee

\subsection{Semisimplicity}

Assuming that $\cA$ is semisimple, in the sense that the bilinear form $g$ is non-degenerate, the forms (\ref{bil}) define a scalar product and a hermitian product.  
We denote by $g^{AB}$ the inverse metric and by $h^{A\bar B}$ the inverse hermitian form,  that is  $g^{AB} g_{BC} = \delta_C^A$ and $h^{A\bar B}h_{\bar BC}=\delta_{C}^{A}$. 
Note that for any algebra element $M  = M^A e_A$, the coordinates can be written as: $M^A = g^{AB}\Tr_{\cA}(e_B M)$ or $M^A = h^{A\bar A} \Tr_{\cA}(e^\dagger_{\bar A} M)$,
whence 
\be \label{coord}
g^{AB} \Tr_{\cA}(e_B M)  e_A = h^{A\bar A}  \Tr_{\cA}(e^\dagger_{\bar A} M) e_A  = M \; .
\ee
Multiplying by $N \in \cA$ and tracing these relations give the following `dipole' identities:
\be \label{dipole g}
g^{AB} \Tr_{\cA} (M e_B ) \Tr_{\cA} (e_A N) = \Tr_{\cA} (MN) 
\ee 
and 
\be \label{dipole h}
h^{A\bar B} \Tr_{\cA} (M^\dagger e_{\bar B}^\dagger ) \Tr_{\cA} (e_A N) = \Tr_{\cA} (M^\dagger N) 
\ee
for all $M, N \in \cA$. 

\subsection{Matrix algebras}

Any associative semisimple algebra is a direct sum of matrix algebras.  The basic example of such an algebra is thus 
the algebra  $\cA = {\rm GL}(V)$ of endomorphisms of a $d$ dimensional vector space $V$. In this case, the trace $\Tr_{\cA}$ is related to the usual trace in $V$
as
\be
\Tr_{\cA}(M) = d\, {\rm tr}_V(M)
\ee
We have in particular that $\Tr_{\cA}(\mathbb{I}) = {\rm dim}(\cA) = d^2 = d \, {\rm tr}_V(\mathbb{I})$, where   $\mathbb{I}$ 
denotes the identity endomorphism in ${\rm GL}(V)$. Let  $\{e_A\}$ be a basis of $\cA$ and $|n\rangle$ be a an orthonormal basis in $V$. 
By applying Equ. (\ref{coord}) for $M = |p\rangle \langle q|$, we obtain:
\bea
 d \, g^{AB}  \langle q | e_B |p \rangle \;  e_{A}  =  |p\rangle \langle q| \Rightarrow
 d \, g^{AB}  \langle q | e_B |p \rangle \; \langle a | e_{A} | b\rangle = \delta_{ap} \delta_{qb} 
\eea 
choosing $b=q $ and summing we obtain
\bea
d \, g^{AB}  \langle a | e_{A}  e_B |p \rangle = d \; \delta_{ap} \Rightarrow  g^{AB} e_{A}  e_B = \; \mathbb{I} \; .
\eea 
We obtain similarly:
\be
\bar{h}^{AB} e_{A}  e_B^{\dagger} = \; \mathbb{I} \; .
\ee
Note that for any unitary operator $U = U^A e_A$ in $\cA$ we have 
\be
 \overline{U^A } h_{AB} U^B =  d\, \mathrm{tr}_V(U^{\dagger}U) = d^2 \; ,
\ee
and for any three algebra elements $M_i = M_i^Ae_A, \; i=1,2,3$, we have 
  \bea
    d \, \mathrm{tr}_V (M_1M_2M_3) = M_1^{A_1} M_2^{A_2} M_3^{A_3} \, C_{A_1A_2A_3} \; .
  \eea

Given the orthonormal basis $|n\rangle$ of $V$, the standard choice of basis $\{e_A\}$ of the algebra are the elements $e_{(mn)}$  labeled by a pair $A = (mn)$ given by
\bea
  \Bigl[ e_{(nm)} \Bigr]_{ab} = \delta_{na} \delta_{mb} \; .
\eea 
In this basis, the tensors and structure constants read:
\bea 
g_{(nm) (kl)} &=&  
   d\, \mathrm{tr}_V (e_{(nm)} e_{(kl)} ) = d \sum_{ab} \Bigl[ e_{(nm)} \Bigr]_{ba}\Bigl[ e_{(kl)} \Bigr]_{ab}  = d \, \delta_{nl}  \delta_{mk}, \crcr
     h_{(nm) (kl)} &=&  
   d\, \mathrm{tr}_V (e^\dagger_{(nm)} e_{(kl)} ) = d\, \sum_{ab} \Bigl[ e_{(nm)} \Bigr]_{ab} 
\Bigl[ e_{(kl)} \Bigr]_{ab} = d \delta_{nk}  \delta_{ml} \crcr
     C_{(nm)(kl)(pq)}     &=  &  
d\, \mathrm{tr}_V \bigl( e_{(nm)}  e_{(kl)}  e_{(pq)}  \bigr) = d \,\sum_{abc}  \delta_{na} \delta_{mc} 
\delta_{kc} \delta_{lb} \delta_{pb} \delta_{qa} = d\, \delta_{mk}  \delta_{lp} \delta_{nq}
\eea


\begin{thebibliography}{100}

  
\bibitem{overview} 
  L.~Freidel,
  ``Group field theory: An Overview,''
  Int.\ J.\ Theor.\ Phys.\  {\bf 44}, 1769 (2005)
  [hep-th/0505016].
 
\bibitem{mm}
  P.~Di Francesco, P.~H.~Ginsparg and J.~Zinn-Justin,
  ``2-D Gravity and random matrices,''
  Phys.\ Rept.\  {\bf 254}, 1 (1995)
  [arXiv:hep-th/9306153].

\bibitem{tenquestions} 
  A.~Baratin and D.~Oriti,
  ``Ten questions on Group Field Theory (and their tentative answers),''
  J.\ Phys.\ Conf.\ Ser.\  {\bf 360}, 012002 (2012)
  [arXiv:1112.3270 [gr-qc]].
  
\bibitem{color1}
  R.~Gurau,
  ``Colored Group Field Theory,''
  Commun.\ Math.\ Phys.\  {\bf 304}, 69 (2011)
  [arXiv:0907.2582 [hep-th]].

  
\bibitem{Raz2} 
  R.~Gurau,
  ``The complete 1/N expansion of colored tensor models in arbitrary dimension,''
  Annales Henri Poincare {\bf 13}, 399 (2012)
  [arXiv:1102.5759 [gr-qc]].
  
\bibitem{melogft} 
  A.~Baratin, S.~Carrozza, D.~Oriti, J.~P.~Ryan and M.~Smerlak,
  ``Melonic phase transition in group field theory,'' to appear in Lett.\ Math.\ Phys.  
  [arXiv:1307.5026 [hep-th]].
 
\bibitem{su2} 
  S.~Carrozza, D.~Oriti and V.~Rivasseau,
  ``Renormalization of an SU(2) Tensorial Group Field Theory in Three Dimensions,''
  Commun. Math. Phys. (2014)
  [arXiv:1303.6772 [hep-th]].
  
\bibitem{LaurentDavid}
   L.~Freidel and D.~Louapre,
  ``Diffeomorphisms and spin foam models,''
  Nucl.\ Phys.\ B {\bf 662}, 279 (2003)
  [arXiv:gr-qc/0212001].
  
  \bibitem{Baratin} 
  A.~Baratin, F.~Girelli and D.~Oriti,
  ``Diffeomorphisms in group field theories,'' 
  Phys.\ Rev.\ D {\bf 83}, 104051 (2011)
  [arXiv:1101.0590 [hep-th]].
  
  \bibitem{Baratin-Oriti} 
  A.~Baratin and D.~Oriti,
  ``Group field theory with non-commutative metric variables,''
  Phys.\ Rev.\ Lett.\  {\bf 105}, 221302 (2010)
  [arXiv:1002.4723 [hep-th]].

\bibitem{Flo} 
  M.~Dupuis, F.~Girelli and E.~R.~Livine,
  ``Spinors and Voros star-product for Group Field Theory: First Contact,''
  Phys.\ Rev.\ D {\bf 86}, 105034 (2012)
  [arXiv:1107.5693 [gr-qc]].
  


\bibitem{Oeckl}
  R.~Oeckl,
  ``Braided quantum field theory,''
  Commun.\ Math.\ Phys.\  {\bf 217}, 451 (2001)
  [arXiv:hep-th/9906225].
  
  \bibitem{FloEtera}  
F.\ Girelli and E.\ R. Livine, ``Field theories with homogenous momentum space'', 
Proceedings of the XXV Max Born Symposium, "The Planck Scale", Wroclaw, Poland, July 2009. [arXiv:0910.3107[hep-th]].

\bibitem{Balachandran}
A. P. Balachandran, A. Pinzul, B. A. Qureshi, ``Twisted PoincarŽ Invariant Quantum Field Theories'', Phys.\ Rev.\  D {\bf 77} 025021 (2008). [arXiv:0708.1779[hep-th]]. 

  
  \bibitem{LivineF} 
  L.~Freidel and E.~R.~Livine,
  ``Effective 3-D quantum gravity and non-commutative quantum field theory,''
  Phys.\ Rev.\ Lett.\  {\bf 96}, 221301 (2006)
  [arXiv:hep-th/0512113].

\bibitem{LivineF1}
  L.~Freidel and E.~R.~Livine,
  ``Ponzano-Regge model revisited III: Feynman diagrams and effective field theory,''
  Class.\ Quant.\ Grav.\  {\bf 23}, 2021 (2006)
  [arXiv:hep-th/0502106].
  
\bibitem{Sasai}
 Y.~Sasai and N.~Sasakura,
  ``Braided quantum field theories and their symmetries,''
  Prog.\ Theor.\ Phys.\  {\bf 118}, 785 (2007)
  [arXiv:0704.0822 [hep-th]].
  
  
\bibitem{Boulatov}
 D.~V.~Boulatov,
  ``A Model of three-dimensional lattice gravity,''
  Mod.\ Phys.\ Lett.\ A {\bf 7}, 1629 (1992)
  [arXiv:hep-th/9202074].
  
\bibitem{color2}
  J.~Ben Geloun, J.~Magnen and V.~Rivasseau,
  ``Bosonic Colored Group Field Theory,''
  Eur.\ Phys.\ J.\ C {\bf 70}, 1119 (2010)
  [arXiv:0911.1719 [hep-th]].

\bibitem{BaisMuller}  
F.\ A.\ Bais and N.\ M.\ Muller, ``Topological field theory and the quantum double of SU(2)'',
Nucl.\ Phys.\ B {\bf 530}, 349 (1998) [arXiv:hep-th/9804130].

\bibitem{PRII}
 L.\ Freidel and D.\  Louapre, `` Ponzano-Regge model revisited II: Equivalence with Chern-Simons'',  arXiv:gr-qc/0410141.
       
\bibitem{Livine-Girelli}
F.~Girelli and E.~Livine, ``A Deformed Poincare Invariance for Group Field Theories''
Class.\ Quantum Grav.\ {\bf 27} 245018 (2010). [arXiv:1001.2919[gr-qc]]. 
  


  
\bibitem{Pezzana}
  M. Pezzana, Sulla struttura topologica delle varieta compatte, Atti Sem. Mat. Fis. Univ. Modena, 23 (1974), 269-277.

\bibitem{universality}
R. Gurau, ``Universality for Random Tensors'', to appear in Annales de l'institut Henri Poincar\' (b), Probability and Statistics  arXiv:1111.0519[math.PR].

 \bibitem{FHK}
M. Fukuma, S. Hosono, H. Kawai, ``Lattice topological field theory in two-dimensions'',
Commun.\ Math.\ Phys. {\bf 161},  157-176 (1994). [arXiv:hep-th/9212154]
   
\bibitem{Turaev}
V.Turaev, ``Dijkgraaf-Witten invariants of surfaces and projective representations of groups'', 
arXiv:0706.0160v2 [math.GT].

\bibitem{Synder}
N. Snyder, ``Mednykh's Formula via Lattice Topological Quantum Field Theories'',  
arXiv:math/0703073v3 [math.QA]

\bibitem{DanieleSylvain}
D. Oriti and S. Carrozza, ``Bounding bubbles: the vertex representation of 3d Group Field Theory and the suppression of pseudo-manifolds''
Phys.\ Rev.\ D {\bf 85}, 044004 (2012).  [arXiv:1104.5158[hep-th]] 
  
\bibitem{Ryan} 
  J.~P.~Ryan,
  ``Tensor models and embedded Riemann surfaces,''
  Phys.\ Rev.\ D {\bf 85}, 024010 (2012)
  [arXiv:1104.5471 [gr-qc]].
   
\bibitem{dipole}
M. Ferri and C. Gagliardi, ``Crystallisation moves'', Pacific journal of mathematics, Vol. 100, No. 1 (1982). 
  
  




\end{thebibliography}
\end{document}